\documentclass[12pt]{article}
\usepackage[latin9]{inputenc}
\usepackage{units}
\usepackage{textcomp}
\usepackage{amsmath}
\usepackage{amssymb}
\usepackage[numbers,square,sort&compress]{natbib}
\usepackage[unicode=true,pdfusetitle,
 bookmarks=true,bookmarksnumbered=false,bookmarksopen=false,
 breaklinks=false,pdfborder={0 0 1},backref=false,colorlinks=false]
 {hyperref}
\hypersetup{
 pdfborderstyle=}

\makeatletter
\numberwithin{equation}{section}

\usepackage{amsfonts}

\makeatother

\begin{document}
\begin{center}
\textbf{\large{}On Unfolded Approach To Off-Shell Supersymmetric Models}{\large\par}
\par\end{center}

\begin{center}
 
\par\end{center}

\begin{center}
\vspace{0.2cm}
 \textbf{N.G.~Misuna}\\
 \vspace{0.5cm}
 \emph{Max-Planck-Institut für Gravitationsphysik (Albert-Einstein-Institut),}\\
 \emph{ Am Mühlenberg 1, 14476, Potsdam, Germany }\\
 \vspace{0.5cm}
 \textit{Tamm Department of Theoretical Physics, Lebedev Physical
Institute,}\\
 \textit{Leninsky prospekt 53, 119991, Moscow, Russia}\\
 
\par\end{center}

\begin{center}
\vspace{0.6cm}
 nikita.misuna@aei.mpg.de \\
 
\par\end{center}

\vspace{0.4cm}

\begin{abstract}
\noindent We construct and analyze unfolded off-shell systems for
chiral and vector supermultiplets using multispinor formalism and
external currents. We find that auxiliary variables of multispinor
formalism allow for the interesting reorganization of the unfolded
off-shell modules: one can unify dynamical and auxiliary scalars in
the Wess-Zumino model, as well as an auxiliary pseudoscalar and a
zero-helicity subsector in the vector supermultiplet, so that the
resulting unfolded equations get simplified, while various constraints
like chirality, electric current conservation and P-oddness of the
pseudoscalar can be formulated entirely in terms of these auxiliary
variables.

\newpage{}

\tableofcontents{} 
\end{abstract}

\section{Introduction}

Unfolded dynamics approach \citep{unf1,unf2,unf4} was designed to
formulate the higher-spin gravity theory within the frame of Vasiliev
equations \citep{vaseq1,vaseq2}. In a nutshell, the unfolded dynamics
approach to a field theory amounts to reformulating it as a set of
first-order non-Lagrangian equations, which are manifestly coordinate-independent
and gauge-invariant, by means of introducing (usually infinite number
of) descendant fields which parametrize all derivatives of the primary
fields. Coordinate-independence and gauge invariance make the unfolded
approach attractive beyond the original higher-spin gravity problem.
In particular, the unfolded approach proposes new interesting possibilities
for constructing manifestly supersymmetric formulations \citep{susnf0,susnf1,susnf2,susnf3},
which is topical e.g. for maximally supersymmetric Yang--Mills theories.

From the point of view of the unfolded approach, global symmetries
(e.g. (super)-Poincaré symmetry) emerge as residual gauge symmetries
of non-dynamical vacuum fields (gravitational field for Poincaré symmetry,
plus gravitino fields for SUSY). Hence consistent inclusion of the
background fields to the unfolded system provides manifest realization
of global symmetries. Moreover, unfolded equations are formulated
in terms of differential forms, that guarantees manifest diffeomorphism
invariance of the whole formalism, which is a crucial feature when
dealing with a theory that contains dynamical gravity (e.g. higher-spin
gravity). This allows, in addition, a straightforward lift of an unfolded
supersymmetric theory from Minkowski space to superspace \citep{susnf0,susnf1}.
And the problem of constructing an off-shell completion for a given
on-shell unfolded system, as was shown in \citep{misuna_1}, amounts
to coupling the on-shell system to external currents, which are then
interpreted as off-shell descendant fields.

In this paper we construct and analyze unfolded formulations for $4d$
off-shell Wess--Zumino and vector supermultiplets, using a multispinor
formalism and proposals of \citep{misuna_1}. Infinite sequences of
unfolded descendants are collected into master-fields as expansions
in auxiliary spinors $Y=(y^{\alpha},\bar{y}^{\dot{\alpha}})$ and
scalar $p$. An advantage of the $4d$ multispinor formalism is that
it automatically imposes tracelessness in derivatives (following directly
from the commutativity of $Y$-spinors) on the space of unfolded descendants,
that effectively puts the system in question on the mass shell. On
the other hand, this does not allow one to simply relax the tracelessness,
which is possible in tensorial unfolded formulation and represents
a standard way to generate unfolded off-shell extensions in that case
\citep{off1,off2,off3,unf4,susnf1}. A way out, proposed in \citep{misuna_1},
consists in introducing, on top of $Y$, an additional variable $p$,
which accounts for the off-shell descendants.

Here we find that this variable allows for a curious reorganization
of the supersymmetric unfolded modules: some dynamical and auxiliary
component fields (in particular, scalars in the Wess--Zumino supermultiplet,
and a pseudoscalar and zero-helicity subsector of the vector field
in the vector supermultiplet) can be combined into united unfolded
modules as even and odd parts in $p$-expansions, which leads to a
significant simplification of the unfolded equations. Moreover, it
turns out that then the various constraints get reformulated as relations
solely in terms of auxiliary $Y-p$ variables: chirality constraint
and a non-canonical dimension of the auxiliary scalar for the Wess--Zumino
model, conservation of the electric current and $P$-oddness of the
auxiliary pseudoscalar for the vector supermultiplet. Given that such
a reorganization does not seem to have any immediate simple interpretation
in the conventional Lagrangian terms, the proposed approach can provide
new tools for dealing with the off-shell supersymmetry problems.

The paper is organized as follows. In Section \ref{SEC_UDA} we briefly
discuss the unfolded dynamics approach and give concrete simple examples
of unfolded systems used later. In Section \ref{SEC_WZ} we present
the unfolded off-shell Wess--Zumino model in spinorial formulation,
alternative to the tensorial formulation of \citep{susnf1}, and show
that it allows one to reduce unfolded off-shell modules to a simpler
form. Section \ref{SEC_VECTOR} is devoted to the unfolding of the
off-shell vector supermultiplet. In Section \ref{SEC_CONCLUSION}
we summarize our results.

\section{Unfolded approach\label{SEC_UDA}}

Unfolded formulation \citep{unf1,unf2,unf4} of the field theory represents
a set of equations of the form 
\begin{equation}
\mathrm{d}W^{A}(x)+G^{A}(W)=0,\label{unf_eq}
\end{equation}
on unfolded fields $W^{A}(x)$, which are differential forms, with
$A$ standing for all indices they carry. Here $\mathrm{d}$ is the
de Rham differential on the space-time (super)manifold $M^{d}$ with
local coordinates $x$ and $G^{A}(W)$ is built from exterior products
of $W$ (wedge symbol will be always implicit). The identity $\mathrm{d}^{2}\equiv0$
entails the following \emph{consistency condition} for an unfolded
system 
\begin{equation}
G^{B}\dfrac{\delta G^{A}}{\delta W^{B}}\equiv0,\label{unf_consist}
\end{equation}
which is of the first importance in the unfolding procedure. Equations
\eqref{unf_eq} are invariant under a set of infinitesimal gauge transformations
\begin{equation}
\delta W^{A}=\mathrm{d}\varepsilon^{A}-\varepsilon^{B}\dfrac{\delta G^{A}}{\delta W^{B}}.\label{unf_gauge_transf}
\end{equation}
Gauge parameter $\varepsilon^{A}(x)$ is a rank-$(r-1)$ form related
to a rank-$r$ unfolded field $W^{A}$. 0-forms do not have their
own gauge symmetries and are transformed only by higher-forms gauge
transformations through the second term in \eqref{unf_gauge_transf}.

The sense of the unfolded system \eqref{unf_eq} can be understood
as follows. A set of unfolded fields $W^{A}(x)$ decomposes into subsets
of primary fields and descendant fields. The unfolded equations express
the descendants in terms of the primaries and their derivatives. At
the same time, they may (implicitly) put some differential combinations
of primaries to zero, which means that the system is on-shell; otherwise
it is off-shell. Concrete simple examples will be given below.

\subsection{Supersymmetric vacuum}

The non-dynamical maximally-symmetric space-time background in the
unfolded approach arises through a 1-form connection $\Omega=\mathrm{d}x^{\underline{a}}\Omega_{\underline{a}}^{A}(x)T_{A}$,
which takes values in Lie algebra of space-time symmetries with generators
$T_{A}$ and obeys zero-curvature condition (which is of the form
\eqref{unf_eq})
\begin{equation}
\mathrm{d}\Omega+\Omega\Omega=0.\label{flat_conn}
\end{equation}
A choice of some particular solution $\Omega_{0}$ breaks the gauge
symmetry
\begin{equation}
\delta\Omega=\mathrm{d}\varepsilon(x)-[\Omega,\varepsilon]
\end{equation}
(square brackets stands for the Lie-algebra commutator) to the residual
global symmetry $\varepsilon_{glob}$ that leaves $\Omega_{0}$ invariant
and is determined by
\begin{equation}
\mathrm{d}\varepsilon_{glob}-[\Omega_{0},\varepsilon_{glob}]=0.\label{glob_symm}
\end{equation}
Because there are no 0-forms in \eqref{flat_conn}, the background
is non-dynamical: it describes no gauge-invariant physical d.o.f.

In the paper we consider $4d$ $\mathcal{N}=1$ super-Poincaré group,
so an appropriate 1-form is a connection
\begin{equation}
\Omega=e^{\alpha\dot{\beta}}P_{\alpha\dot{\beta}}+i\omega^{\alpha\beta}M_{\alpha\beta}+i\bar{\omega}^{\dot{\alpha}\dot{\beta}}\bar{M}_{\dot{\alpha}\dot{\beta}}+\psi^{\alpha}Q_{\alpha}+\bar{\psi}^{\dot{\alpha}}\bar{Q}_{\dot{\alpha}},
\end{equation}
where $P_{\alpha\dot{\alpha}}$, $M_{\alpha\beta}$, $Q_{\alpha}$
are generators of translations, Lorentz transformations and supercharges,
while $e^{\alpha\dot{\beta}}$, $\Omega^{\alpha\beta}$ , $\psi^{\alpha}$
are 1-forms of vierbein, Lorentz connection and gravitino ($\alpha$
and $\dot{\alpha}$ are $sl(2,\mathbb{C})$-spinor indices). For this
$\Omega$ \eqref{flat_conn} gives, accounting for commutation relations
of generators,
\begin{align}
 & \mathrm{d}e^{\alpha\dot{\beta}}+\omega^{\alpha}\text{}_{\gamma}e^{\gamma\dot{\beta}}+\bar{\omega}^{\dot{\beta}}\text{}_{\dot{\gamma}}e^{\alpha\dot{\gamma}}-\psi^{\alpha}\bar{\psi}^{\dot{\beta}}=0,\\
 & \mathrm{d}\omega^{\alpha\beta}+\omega^{\alpha}\text{}_{\gamma}\omega^{\gamma\beta}=0,\qquad\mathrm{d}\bar{\omega}^{\dot{\alpha}\dot{\beta}}+\bar{\omega}^{\dot{\alpha}}\text{}_{\dot{\gamma}}\bar{\omega}^{\dot{\gamma}\dot{\beta}}=0,\\
 & \mathrm{d}\psi^{\alpha}+\omega^{\alpha}\text{}_{\gamma}\psi^{\gamma}=0,\qquad\mathrm{d}\bar{\psi}^{\dot{\alpha}}+\bar{\omega}^{\dot{\alpha}}\text{}_{\dot{\gamma}}\bar{\psi}^{\dot{\gamma}}=0.
\end{align}
Fixing some particular solution to these equations reduces initial
supergravity gauge symmetry \eqref{unf_gauge_transf} to a global
supersymmetry \eqref{glob_symm} and determines supertransformation
rules for all fields coupled to this background.

\subsection{Scalar field\label{subsec:Scalar-field}}

The simplest dynamical system is a free scalar field. An appropriate
set of unfolded fields to describe it includes all multispinor fields
of the type $(\frac{n}{2},\frac{n}{2})$, i.e. $C_{\alpha(n),\dot{\alpha}(n)}(x)$
for all $n$ (we make use of the condensed notations in the paper,
denoting a set of $n$ symmetrized indices $A_{1}...A_{n}$ as $A(n)$)
\citep{unf2}. All these fields can be collected into a single unfolded
module by means of the auxiliary commuting spinors $Y=(y^{\alpha},\bar{y}^{\dot{\alpha}})$,
$\alpha,\dot{\alpha}=\overline{1,2}$ . Spinor indices are raised
and lowered by antisymmetric metrics
\begin{equation}
\epsilon_{\alpha\beta}=\epsilon_{\dot{\alpha}\dot{\beta}}=\left(\begin{array}{cc}
0 & 1\\
-1 & 0
\end{array}\right),\quad\epsilon^{\alpha\beta}=\epsilon^{\dot{\alpha}\dot{\beta}}=\left(\begin{array}{cc}
0 & 1\\
-1 & 0
\end{array}\right)
\end{equation}
as
\begin{equation}
v^{\alpha}=\epsilon^{\alpha\beta}v_{\beta},\quad v_{\alpha}=\epsilon_{\beta\alpha}v^{\beta},\quad\bar{v}^{\dot{\alpha}}=\epsilon^{\dot{\alpha}\dot{\beta}}\bar{v}_{\dot{\beta}},\quad\bar{v}_{\dot{\alpha}}=\epsilon_{\dot{\beta}\dot{\alpha}}\bar{v}^{\dot{\beta}},
\end{equation}
so because of commutativity
\begin{equation}
y^{\alpha}y^{\beta}\epsilon_{\alpha\beta}=0,\quad\bar{y}^{\dot{\alpha}}\bar{y}^{\dot{\beta}}\epsilon_{\dot{\alpha}\dot{\beta}}=0.
\end{equation}
A single unfolded scalar module is then defined as
\begin{equation}
C(Y|x)=\sum_{n=0}^{\infty}\frac{1}{(n!)^{2}}C_{\alpha(n),\dot{\alpha}(n)}(y^{\alpha})^{n}(\bar{y}^{\dot{\alpha}})^{n},
\end{equation}
and an unfolded equation for $C$ is \citep{vaseq2}
\begin{equation}
DC+ie\partial\bar{\partial}C=0.\label{unf_scal}
\end{equation}
Here
\begin{equation}
D=\mathrm{d}+\omega^{\alpha\beta}y_{\alpha}\partial_{\beta}+\bar{\omega}^{\dot{\alpha}\dot{\beta}}\bar{y}_{\dot{\alpha}}\bar{\partial}_{\dot{\beta}}
\end{equation}
is a 1-form of a Lorentz-covariant derivative, where $Y$-derivatives
are introduced as
\begin{equation}
\partial_{\alpha}y^{\beta}=\delta_{\alpha}\text{}^{\beta},\quad\bar{\partial}_{\dot{\alpha}}\bar{y}^{\dot{\beta}}=\delta_{\dot{\alpha}}\text{}^{\dot{\beta}},
\end{equation}
and for brevity we denote
\begin{equation}
ey\bar{y}=e^{\alpha\dot{\beta}}y_{\alpha}\bar{y}_{\dot{\beta}},\quad e\partial\bar{\partial}=e^{\alpha\dot{\beta}}\partial_{\alpha}\bar{\partial}_{\dot{\beta}},\quad ey\bar{\partial}=e^{\alpha\dot{\beta}}y_{\alpha}\bar{\partial}_{\dot{\beta}},\quad e\bar{y}\partial=e^{\alpha\dot{\beta}}\bar{y}_{\dot{\beta}}\partial_{\alpha}.
\end{equation}
Representing $D$ as
\begin{equation}
D=e^{\alpha\dot{\beta}}D_{\alpha\dot{\beta}}\label{D=00003DeD}
\end{equation}
one gets from \eqref{unf_scal} 
\begin{align}
 & C_{\alpha(n),\dot{\alpha}(n)}(x)=(iD_{\alpha\dot{\alpha}})^{n}C(0|x),\label{scal_desc}\\
 & \square C(0|x)=0.\label{KG}
\end{align}
Thus $C_{\alpha(n),\dot{\alpha}(n)}$ are descendant fields, forming
a tower of symmetrized traceless derivatives of the on-shell primary
field $C(0|x)$ which is subjected to the massless Klein--Gordon
equation.

To construct an off-shell completion for the unfolded on-shell system
one has to couple it to external currents, as explained in \citep{misuna_1}.
These currents, sourcing r.h.s. of e.o.m. (Klein--Gordon equation
in this case), further can be interpreted as off-shell descendants
so that former e.o.m. turn to constraints expressing these descendants
in terms of primaries.

As shown in \citep{misuna_1}, all off-shell unfolded scalar fields
can be collected in a single module by a simple introduction of one
more auxiliary scalar variable $p$, so that an off-shell unfolded
scalar module is
\begin{equation}
C(Y|p|x)=\sum_{M,n=0}^{\infty}\frac{1}{(2M)!(n!)^{2}}C_{\alpha(n),\dot{\alpha}(n)}^{(M)}p^{2M}(y^{\alpha})^{n}(\bar{y}^{\dot{\alpha}})^{n}.\label{scal_off_module}
\end{equation}
Here we used expansion in even powers of $p$, differently from \citep{misuna_1},
what for a separately taken scalar field makes no difference, but
is better suited for the supersymmetric extensions, as we will see
below.

An unfolded equation for an off-shell scalar field is
\begin{equation}
DC+ie\partial\bar{\partial}C+iey\bar{y}\frac{\partial_{p}^{2}}{(\varsigma+1)(\varsigma+2)}C=0,\label{off-shell_scalar}
\end{equation}
where
\begin{align}
 & \partial_{p}=\frac{\partial}{\partial p},\\
 & \varsigma=\frac{(N+\bar{N})}{2},\\
 & N=y^{\alpha}\partial_{\alpha},\qquad\bar{N}=\bar{y}^{\dot{\alpha}}\bar{\partial}_{\dot{\alpha}}.
\end{align}
Now instead of \eqref{scal_desc} one has 
\begin{equation}
C_{\alpha(n),\dot{\alpha}(n)}^{(M)}(x)=\square^{M}(iD_{\alpha\dot{\alpha}})^{n}C(0|x),\label{scal_desc_off}
\end{equation}
and instead of \eqref{KG} 
\begin{equation}
\square C^{(0)}(x)=C^{(1)}(x).\label{KG_source}
\end{equation}
So \eqref{off-shell_scalar} indeed corresponds to a scalar field
$C^{(0)}$ coupled to an external current $C^{(1)}$. Or, as we take
it, it corresponds to an unfolded off-shell scalar $C^{(0)}$ with
$C^{(1)}$ being one of its descendants determined by \eqref{KG_source}.
And as one can see from the comparison of \eqref{off-shell_scalar}
and \eqref{scal_desc_off}, an expansion of $C$ in powers of $p^{2}$
is equivalent to an expansion in powers of boxes of the primary scalar.

\subsection{Vector gauge field}

Maxwell field in the unfolded language corresponds to a 1-form of
the vector potential $A=A_{\underline{m}}(x)\mathrm{d}x^{\underline{m}}$
and two conjugate 0-form modules
\begin{equation}
F(Y|x)=\sum_{n=0}^{\infty}\frac{1}{n!(n+2)!}F_{\alpha(n+2),\dot{\alpha}(n)}(y^{\alpha})^{n+2}(\bar{y}^{\dot{\alpha}})^{n},\quad\bar{F}(Y|x)=\sum_{n=0}^{\infty}\frac{1}{n!(n+2)!}\bar{F}_{\alpha(n),\dot{\alpha}(n+2)}(y^{\alpha})^{n}(\bar{y}^{\dot{\alpha}})^{n+2}.\label{vec_mod}
\end{equation}
Unfolded equations are \citep{vaseq2}
\begin{align}
 & \mathrm{d}A=\frac{i}{4}E\partial\partial F+\frac{i}{4}\bar{E\partial\partial}\bar{F},|_{Y=0}\label{vec_eq_1}\\
 & DF+ie\partial\bar{\partial}F=0,\\
 & D\bar{F}+ie\partial\bar{\partial}\bar{F}=0,\label{vec_eq_2}
\end{align}
from which one finds
\begin{align}
 & F_{\alpha\alpha}=\frac{i}{2}D_{\alpha}{}^{\dot{\beta}}(\sigma^{\underline{m}})_{\alpha\dot{\beta}}A_{\underline{m}},\qquad\bar{F}_{\dot{\alpha}\dot{\alpha}}=\frac{i}{2}D^{\beta}{}_{\dot{\alpha}}(\sigma^{\underline{m}})_{\beta\dot{\alpha}}A_{\underline{m}}.\label{F_A}\\
 & F_{\alpha(n+2),\dot{\alpha}(n)}=(iD_{\alpha\dot{\alpha}})^{n}F_{\alpha\alpha},\qquad\bar{F}_{\alpha(n),\dot{\alpha}(n+2)}=(iD_{\alpha\dot{\alpha}})^{n}\bar{F}_{\dot{\alpha}\dot{\alpha}}.\label{F_desc}\\
 & D^{\beta}{}_{\dot{\alpha}}F_{\beta\alpha}=0,\qquad D_{\alpha}{}^{\dot{\beta}}\bar{F}_{\dot{\beta}\dot{\alpha}}=0.\label{Maxwell_eq}
\end{align}
Thus, $\bar{F}$ and $F$ encode selfdual $\bar{F}_{\dot{\alpha}\dot{\alpha}}$
and anti-selfdual $F_{\alpha\alpha}$ parts of the Maxwell tensor
and all their on-shell derivatives \eqref{F_desc}. The Maxwell tensor
is built from the vector potential $A_{\underline{m}}$ according
to \eqref{F_A} and obeys Maxwell equations \eqref{Maxwell_eq}. From
the general formula \eqref{unf_gauge_transf} one restores a conventional
gauge symmetry of electrodynamics
\begin{equation}
\delta A=\mathrm{d}\epsilon(x),\qquad\delta F=0,\qquad\delta\bar{F}=0.
\end{equation}

An off-shell vector field arises through coupling of \eqref{vec_eq_1}-\eqref{vec_eq_2}
to an external conserved electric current. As shown in \citep{misuna_1},
an unfolded electric current module $J(Y|p|x)$ is structured as follows
\begin{align}
 & J(Y|p|x)=J^{0}+J^{+}+J^{-},\\
 & J^{0}=\sum_{M,n=0}^{\infty}\frac{1}{(2M)!(n+1)!^{2}}J_{\alpha(n+1),\dot{\alpha}(n+1)}^{(M)}p^{2M}(y^{\alpha})^{n+1}(\bar{y}^{\dot{\alpha}})^{n+1},\\
 & J^{+}=\sum_{M,n=0}^{\infty}\frac{1}{(2M)!(n+2)!n!}J_{\alpha(n+2),\dot{\alpha}(n)}^{(M)}p^{2M}(y^{\alpha})^{n+2}(\bar{y}^{\dot{\alpha}})^{n},\\
 & J^{-}=\sum_{M,n=0}^{\infty}\frac{1}{(2M)!(n+2)!n!}J_{\alpha(n),\dot{\alpha}(n+2)}^{(M)}p^{2M}(y^{\alpha})^{n}(\bar{y}^{\dot{\alpha}})^{n+2},
\end{align}
and an off-shell system for a free vector gauge field is
\begin{align}
 & \mathrm{d}A=\frac{i}{4}E\partial\partial F+\frac{i}{4}\bar{E\partial\partial}\bar{F},|_{Y=0}\label{vec_eq_off_1}\\
 & DF+ie\partial\bar{\partial}F+iey\bar{y}\frac{1}{(\varsigma+1)(\varsigma+2)}J^{+}+ey\bar{\partial}\frac{2}{(\varsigma+1)(\varsigma+2)}J^{0}=0,\label{DF_J1}\\
 & D\bar{F}+ie\partial\bar{\partial}\bar{F}+iey\bar{y}\frac{1}{(\varsigma+1)(\varsigma+2)}J^{-}+e\partial\bar{y}\frac{2}{(\varsigma+1)(\varsigma+2)}J^{0}=0,\label{DF_J2}\\
 & DJ+ie\partial\bar{\partial}J+iey\bar{y}\frac{\varsigma(\varsigma+3)}{(N+1)(N+2)(\bar{N}+1)(\bar{N}+2)}\partial_{p}^{2}J+\nonumber \\
 & +ey\bar{\partial}\frac{1}{(N+1)(N+2)}(J^{-}+2\partial_{p}^{2}J^{0})+e\partial\bar{y}\frac{1}{(\bar{N}+1)(\bar{N}+2)}(J^{+}+2\partial_{p}^{2}J^{0})=0.\label{vec_eq_off_2}
\end{align}
Equation \eqref{vec_eq_off_2} determines all unfolded fields in $J$
in terms of the primary electric current $J_{\alpha,\dot{\alpha}}^{(0)}(x)$
and imposes a conservation condition
\begin{equation}
D_{\alpha\dot{\alpha}}J^{(0)\alpha,\dot{\alpha}}(x)=0
\end{equation}
(this can be deduced by using \eqref{D=00003DeD} and acting on \eqref{vec_eq_off_2}
with $\partial^{\alpha}\bar{\partial}^{\dot{\alpha}}\frac{\delta}{\delta e^{\alpha\dot{\alpha}}}$,
followed by putting $Y=0$, $p=0$). Then \eqref{DF_J1}-\eqref{DF_J2}
describe gluing of the current module $J$ to the on-shell vector
module, thus providing and off-shell completion for the latter.

\subsection{Spinor field}

In \citep{misuna_1} only bosonic fields were considered. Here we
present an unfolded description of an off-shell spin-$\nicefrac{1}{2}$
field which reveals an important peculiarity of fermions.

Weyl $(\frac{1}{2},0)$-spinor is described by an unfolded module
\begin{equation}
\chi(Y|x)=\sum_{n=0}^{\infty}\frac{1}{n!(n+1)!}\chi_{\alpha(n+1),\dot{\alpha}(n)}(x)(y^{\alpha})^{n+1}(\bar{y}^{\dot{\alpha}})^{n}
\end{equation}
supported by an unfolded equation \citep{vaseq2}
\begin{equation}
D\chi+ie\partial\bar{\partial}\chi=0.\label{Weyl_eq}
\end{equation}
Analogously to the scalar case, higher-rank multispinors are expressed
through a primary field $\chi_{\alpha}(x)$ subjected to the Weyl
equation
\begin{align}
 & \chi_{\alpha(n+1),\dot{\alpha}(n)}=(iD_{\alpha\dot{\alpha}})^{n}\chi_{\alpha},\\
 & D_{\alpha\dot{\beta}}\chi^{\alpha}=0.
\end{align}
To put this system off-shell, one has to couple it to an external
current, which in this case is an unconstrained (i.e. also off-shell)
$(0,\frac{1}{2})$-spinor field. An appropriate unfolded off-shell
module turns out to be
\begin{align}
 & \chi(Y|p|x)=\chi^{+}+\chi^{-},\label{x_mod_+-}\\
 & \chi^{+}=\sum_{M,n=0}^{\infty}\frac{1}{(2M)!n!(n+1)!}\chi_{\alpha(n+1),\dot{\alpha}(n)}^{(M)}p^{2M}(y^{\alpha})^{n+1}(\bar{y}^{\dot{\alpha}})^{n},\label{x_mod_+}\\
 & \chi^{-}=\sum_{M,n=0}^{\infty}\frac{1}{(2M)!n!(n+1)!}\chi_{\alpha(n),\dot{\alpha}(n+1)}^{(M)}p^{2M+1}(y^{\alpha})^{n}(\bar{y}^{\dot{\alpha}})^{n+1},\label{x_mod_-}
\end{align}
and unfolded equations are
\begin{equation}
D\chi+ie\partial\bar{\partial}\chi+iey\bar{y}\frac{\partial_{p}^{2}}{(\varsigma+\frac{3}{2})^{2}}\chi+ey\bar{\partial}\frac{\partial_{p}}{(\varsigma+\frac{1}{2})(\varsigma+\frac{3}{2})}\Pi^{-}\chi+e\partial\bar{y}\frac{\partial_{p}}{(\varsigma+\frac{1}{2})(\varsigma+\frac{3}{2})}\Pi^{+}\chi=0,\label{x_off_eq}
\end{equation}
where we introduced projectors on positive and negative helicity (identified
with the difference between the number of undotted and dotted spinors)
as
\begin{align}
 & f_{m,n}=f_{\alpha(m),\dot{\beta}(n)}(y^{\alpha})^{m}(\bar{y}^{\dot{\beta}})^{n},\nonumber \\
 & \Pi^{+}f_{m,n}=\begin{cases}
f_{m,n}, & m\geq n\\
0, & m<n
\end{cases} &  & \Pi^{-}f_{m,n}=\begin{cases}
f_{m,n}, & m\leq n\\
0, & m>n
\end{cases}.
\end{align}
As one sees from \eqref{x_off_eq}, unlike unfolded off-shell bosons,
off-shell fermions require all powers of $p$ to be presented in the
module. And helicities of unfolded fields are related to their $p$-parity,
as follows from \eqref{x_mod_+}-\eqref{x_mod_-}: positive helicities
belong to the $p$-even sector, while negative ones -- to the $p$-odd.

To get an unfolded description for an off-shell Weyl $(0,\frac{1}{2})$-type
spinor $\bar{\zeta}_{\dot{\alpha}}$ one just needs to flip the relation
between helicity and $p$-parity: 
\begin{align}
 & \bar{\zeta}(Y|p|x)=\bar{\zeta}^{-}+\bar{\zeta}^{+},\\
 & \bar{\zeta}^{-}=\sum_{M,n=0}^{\infty}\frac{1}{(2M)!n!(n+1)!}\bar{\zeta}_{\alpha(n),\dot{\alpha}(n+1)}^{(M)}p^{2M}(y^{\alpha})^{n}(\bar{y}^{\dot{\alpha}})^{n+1},\\
 & \bar{\zeta}^{+}=\sum_{M,n=0}^{\infty}\frac{1}{(2M)!n!(n+1)!}\bar{\zeta}_{\alpha(n+1),\dot{\alpha}(n)}^{(M)}p^{2M+1}(y^{\alpha})^{n+1}(\bar{y}^{\dot{\alpha}})^{n}.
\end{align}
while the unfolded equation remains the same 
\begin{equation}
D\bar{\zeta}+ie\partial\bar{\partial}\bar{\zeta}+iey\bar{y}\frac{\partial_{p}^{2}}{(\varsigma+\frac{3}{2})^{2}}\bar{\zeta}+ey\bar{\partial}\frac{\partial_{p}}{(\varsigma+\frac{1}{2})(\varsigma+\frac{3}{2})}\Pi^{-}\bar{\zeta}+e\partial\bar{y}\frac{\partial_{p}}{(\varsigma+\frac{1}{2})(\varsigma+\frac{3}{2})}\Pi^{+}\bar{\zeta}=0.
\end{equation}
Finally, to describe an off-shell Dirac spinor one simply unites two
modules $\chi$ and $\bar{\zeta}$ into an unfolded off-shell Dirac
module 
\begin{equation}
\varXi(Y|p|x)=\chi+\bar{\zeta},
\end{equation}
and the unfolded equation remains the same once again
\begin{equation}
D\varXi+ie\partial\bar{\partial}\varXi+iey\bar{y}\frac{\partial_{p}^{2}}{(\varsigma+\frac{3}{2})^{2}}\varXi+ey\bar{\partial}\frac{\partial_{p}}{(\varsigma+\frac{1}{2})(\varsigma+\frac{3}{2})}\Pi^{-}\varXi+e\partial\bar{y}\frac{\partial_{p}}{(\varsigma+\frac{1}{2})(\varsigma+\frac{3}{2})}\Pi^{+}\varXi=0.\label{Dirac_off_eq}
\end{equation}
But now there is no relation between helicity and $p$-parity: both
helicities have all terms in $p$-expansion.

Thus, for off-shell spinors of all types one has one and the same
unfolded equation, but the structure of the unfolded module does depend
on the type. Introducing projectors on $p$-even and $p$-odd parts
as
\begin{equation}
\Pi^{e}f(p)=\frac{f(p)+f(-p)}{2},\qquad\Pi^{o}f(p)=\frac{f(p)-f(-p)}{2},
\end{equation}
one can formulate this dependence as a constraint on the mutual $p-Y$
dependence
\begin{align}
\textrm{left Weyl:} & \quad\Pi^{e}=\Pi^{+},\label{left_Weyl}\\
\textrm{right Weyl:} & \quad\Pi^{e}=\Pi^{-},\label{right_Weyl}\\
\textrm{Dirac:} & \quad\textrm{no constraints}.\label{Dirac}
\end{align}

\section{Unfolded Wess-Zumino model revisited\label{SEC_WZ}}

Unfolded formulation of the on-shell Wess-Zumino model in terms of
symmetric Lorentz-tensors playing the role of unfolded fields was
built in \citep{susnf0}.
\begin{align}
 & DC^{a(k)}+e_{b}C^{a(k)b}-\sqrt{2}\psi^{\alpha}\chi_{\alpha}^{a(k)}=0,\\
 & D\chi_{\alpha}^{a(k)}+e_{b}\chi_{\alpha}^{a(k)b}-i\sqrt{2}(\sigma_{b})_{\alpha\dot{\beta}}\bar{\psi}^{\dot{\beta}}C^{a(k)b}=0\\
 & C^{a(k-2)b}{}_{b}=0,\qquad\chi_{\alpha}^{a(k-2)b}{}_{b}=0,\qquad(\bar{\sigma}_{b})^{\dot{\alpha}\beta}\chi_{\beta}^{a(k)b}=0.
\end{align}
All tensors are traceless and spinor fields are subjected to $\sigma$-transversality
condition. Off-shell extension arises via relaxing tracelessness and
$\sigma$-transversality conditions and requires introducing a series
of symmetric tensor fields $F^{a(k)}$, which correspond to an auxiliary
component scalar of the chiral supermultiplet \citep{susnf1}. Resulting
off-shell system is
\begin{align}
 & DC^{a(k)}+e_{b}C^{a(k)b}-\sqrt{2}\psi^{\alpha}\chi_{\alpha}^{a(k)}=0,\label{WZ_eq_tens_1}\\
 & D\chi_{\alpha}^{a(k)}+e_{b}\chi_{\alpha}^{a(k)b}-i\sqrt{2}(\sigma_{b})_{\alpha\dot{\beta}}\bar{\psi}^{\dot{\beta}}C^{a(k)b}-\sqrt{2}\psi_{\alpha}F^{a(k)}=0,\\
 & DF^{a(k)}+e_{b}F^{a(k)b}-i\sqrt{2}\bar{\psi}_{\dot{\alpha}}(\bar{\sigma}_{b})^{\dot{\alpha}\beta}\chi_{\beta}^{a(k)b}=0.\label{WZ_eq_tens_3}
\end{align}
Translating this to the multispinor language of this paper, symmetric
traceless tensor $T^{a(k)}$ corresponds to a multispinor $T_{\alpha(k),\dot{\alpha}(k)}$,
while $\sigma$-transverse $\chi_{\alpha}^{a(k)}$ corresponds to
$\chi_{\alpha(k+1),\dot{\alpha}(k)}$. Then $p^{2M}$-terms of \eqref{scal_off_module}
and \eqref{x_mod_+} correspond to the traces of the Lorentz-tensors,
while $p^{2M+1}$-terms from \eqref{x_mod_-} correspond to $\sigma$-longitudinal
contributions of the form $(\bar{\sigma}_{b})^{\dot{\alpha}\beta}\chi_{\beta}^{a(k)b}$.

In order to build a multispinor equivalent of the system \eqref{WZ_eq_tens_1}-\eqref{WZ_eq_tens_3},
one supplements off-shell scalar \eqref{off-shell_scalar} and off-shell
spinor \eqref{x_off_eq} with the unfolded auxiliary scalar
\begin{align}
 & F(Y|p|x)=\sum_{M,n=0}^{\infty}\frac{1}{(2M)!(n!)^{2}}F_{\alpha(n),\dot{\alpha}(n)}^{(M)}p^{2M}(y^{\alpha})^{n}(\bar{y}^{\dot{\alpha}})^{n},\\
 & DF+ie\partial\bar{\partial}F+iey\bar{y}\frac{\partial_{p}^{2}}{(\varsigma+1)(\varsigma+2)}F=0\label{F_eq}
\end{align}
and deforms \eqref{off-shell_scalar}, \eqref{x_off_eq}, \eqref{F_eq}
with $\psi$-dependent terms mixing component $C$, $\chi$ and $F$
in a consistent way. We will not describe here the procedure of this
(and other presented in the paper) unfolding, which is quite technical,
tedious and lengthy. Detailed examples of constructing unfolded systems
can be found e.g. in \citep{misuna_1,misuna_2}. In a nutshell, one
has to write down the most general suitable Ansatz for unfolded equations
with arbitrary $(p|Y)$-dependent coefficients and then fix them by
imposing consistency condition \eqref{unf_consist}.

For the off-shell chiral supermultiplet in question, consistent unfolded
equations turn to be
\begin{align}
 & DC+ie\partial\bar{\partial}C+iey\bar{y}\frac{\partial_{p}^{2}}{(\varsigma+1)(\varsigma+2)}C+\psi\partial\Pi^{+}\chi+i\psi y\frac{\partial_{p}}{(\varsigma+\frac{3}{2})}\Pi^{-}\chi=0,\label{WZ_eq_1}\\
 & D\chi+ie\partial\bar{\partial}\chi+iey\bar{y}\frac{\partial_{p}^{2}}{(\varsigma+\frac{3}{2})^{2}}\chi+ey\bar{\partial}\frac{\partial_{p}}{(\varsigma+\frac{1}{2})(\varsigma+\frac{3}{2})}\Pi^{-}\chi+e\partial\bar{y}\frac{\partial_{p}}{(\varsigma+\frac{1}{2})(\varsigma+\frac{3}{2})}\Pi^{+}\chi-\nonumber \\
 & -i\bar{\psi\partial}C-\bar{\psi y}\frac{\partial_{p}}{(\varsigma+1)}C+i\psi y\frac{1}{(\varsigma+1)}F-p\frac{1}{p\partial_{p}+1}\psi\partial F=0.\\
 & DF+ie\partial\bar{\partial}F+iey\bar{y}\frac{\partial_{p}^{2}}{(\varsigma+1)(\varsigma+2)}F+i\bar{\psi\partial}\partial_{p}\Pi^{-}\chi-\bar{\psi y}\frac{\partial_{p}^{2}}{(\varsigma+\frac{3}{2})}\Pi^{+}\chi=0.\label{WZ_eq_3}
\end{align}
This multispinor formulation allows for a new interesting possibility,
elusive in the tensor form \eqref{WZ_eq_tens_1}-\eqref{WZ_eq_tens_3}.
Namely, it is possible to naturally combine the dynamical and auxiliary
scalar modules $C$ and $F$ into a single one. To this end one defines
a combined scalar module
\begin{equation}
\Phi=C+i\partial_{p}F,
\end{equation}
which thus contains both even (former $C$) and odd (former $F)$
terms in $p$-expansion
\begin{equation}
\Phi(Y|p|x)=\sum_{M,n=0}^{\infty}\frac{1}{M!(n!)^{2}}\Phi_{\alpha(n),\dot{\alpha}(n)}^{(M)}p^{M}(y^{\alpha})^{n}(\bar{y}^{\dot{\alpha}})^{n}.
\end{equation}
Note that this incorporation correctly reproduces a scaling dimension
of the auxiliary scalar, associated with $p$-odd part of $\Phi$:
the scaling dimension of $p$ is
\begin{equation}
\Delta_{p}=1,
\end{equation}
as one can see e.g. from \eqref{scal_off_module} and \eqref{scal_desc_off},
hence
\begin{equation}
\Delta_{F}=\Delta_{C}+1.
\end{equation}
In terms of $\Phi$, \eqref{WZ_eq_1}-\eqref{WZ_eq_3} turns to
\begin{align}
 & D\Phi+ie\partial\bar{\partial}\Phi+iey\bar{y}\frac{\partial_{p}^{2}}{(\varsigma+1)(\varsigma+2)}\Phi+(\psi\partial\Pi^{+}+\bar{\psi\partial}\Pi^{-}+i\psi y\frac{\partial_{p}}{(\varsigma+\frac{3}{2})}\Pi^{-}+i\bar{\psi y}\frac{\partial_{p}}{(\varsigma+\frac{3}{2})}\Pi^{+})\chi=0,\label{WZ_compact_1}\\
 & D\chi+ie\partial\bar{\partial}\chi+iey\bar{y}\frac{\partial_{p}^{2}}{(\varsigma+\frac{3}{2})^{2}}\chi+ey\bar{\partial}\frac{\partial_{p}}{(\varsigma+\frac{1}{2})(\varsigma+\frac{3}{2})}\Pi^{-}\chi+e\partial\bar{y}\frac{\partial_{p}}{(\varsigma+\frac{1}{2})(\varsigma+\frac{3}{2})}\Pi^{+}\chi-\nonumber \\
 & -(i\psi\partial+i\bar{\psi\partial}+\bar{\psi y}\frac{\partial_{p}}{(\varsigma+1)}+\psi y\frac{\partial_{p}}{(\varsigma+1)})\Phi=0.\label{WZ_compact_2}
\end{align}
A curious feature of the system \eqref{WZ_compact_1}-\eqref{WZ_compact_2}
is that naively it looks real, though describing a chiral supermultiplet.
The point is that it is consistent, i.e. satisfying \eqref{unf_consist},
\emph{only} if $\chi$ is a Weyl module, not a Dirac (or Majorana)
one. But as discussed above, unfolded equations for Dirac and both
types of Weyl spinors look completely the same. The difference is
in the structure of the unfolded modules, expressed in \eqref{left_Weyl}-\eqref{Dirac}.
Thus, \eqref{WZ_compact_1}-\eqref{WZ_compact_2} supplemented by
the \emph{chirality constraint}
\begin{equation}
\Pi^{e}\chi=\Pi^{+}\chi,
\end{equation}
determines an unfolded off-shell chiral supermultiplet. And the same
unfolded system \eqref{WZ_compact_1}-\eqref{WZ_compact_2}, but supplemented
with the opposite \emph{anti-chirality constraint} 
\begin{equation}
\Pi^{e}\chi=\Pi^{-}\chi,
\end{equation}
determines an unfolded off-shell anti-chiral supermultiplet.

This ``degeneracy'' of the unfolded equations \eqref{WZ_compact_1}-\eqref{WZ_compact_2}
for chiral and anti-chiral supermultiplets (i.e. that they look completely
the same) becomes manifest only after unifying two scalars of the
Wess--Zumino model into a single module $\Phi$ and is not seen in
the ``standard'' unfolded formulation \eqref{WZ_eq_tens_1}-\eqref{WZ_eq_tens_3}
of \citep{susnf1} or its spinorial version \eqref{WZ_eq_1}-\eqref{WZ_eq_3},
where two supermultiplets are related by non-invariant (i.e. changing
the form of equations) complex conjugation. The main ingredient is
$p$-variable, which allows one to unite two bosonic fields: in a
non-supersymmetric situation bosonic modules depend only on $p^{2}$
which, as is seen e.g. from \eqref{scal_off_module}, \eqref{scal_desc_off},
encodes descendants containing kinetic operators acting on the primary
field; on the other hand, fermions depend on all powers of $p$, because
their kinetic operators are of the first order and change the type
of the spinors. In non-manifestly-supersymmetric reduction of \eqref{WZ_compact_1}-\eqref{WZ_compact_2}
which arises from putting all gravitino 1-forms $\psi$ and $\bar{\psi}$
to zero, $p$-odd and $p$-even terms in $\Phi$ become completely
disentangled -- $\Phi$ divides into independent modules $C$ and
$F$, and the question of their relative $p$-parity becomes inessential.
But in the supersymmetric system non-zero $\psi$ and $\bar{\psi}$
non-trivially intertwine $p$-odd and $p$-even parts of $\Phi$.

Let us also stress that the unification of two scalars and the resulting
degeneracy of the unfolded equations is not easy to explain in terms
of the standard Lagrangian formulation. The reason behind this is
that, as mentioned at the end of Subsection \ref{subsec:Scalar-field},
$p^{2}$ in some sense is conjugate to the wave operator, so that
odd $p$-powers of the bosonic field $\Phi$ do not allow a simple
interpretation.

\section{Unfolded vector supermultiplet\label{SEC_VECTOR}}

In this Section we are about to build and analyze an unfolded system
of an off-shell vector supermultiplet. This is accomplished in several
stages. First, we formulate an on-shell system; then we find an unfolded
description for a supersymmetric source for a vector system, which
is a linear multiplet; finally, we couple a linear multiplet to the
vector system thus arriving at an unfolded off-shell vector supermultiplet.

On-shell vector supermultiplet contains Maxwell field and gaugino
being Majorana spinor. So one has to take \eqref{vec_eq_1}-\eqref{vec_eq_2},
\eqref{Weyl_eq}, add possible terms with $\psi_{\alpha}$, $\bar{\psi}_{\dot{\alpha}}$
mixing the spinor and the vector, and then solve for the consistency
condition \eqref{unf_consist}. This brings to 
\begin{align}
 & \mathrm{d}A=\frac{i}{4}E\partial\partial F+\frac{i}{4}\bar{E\partial\partial}\bar{F}+\frac{1}{2}e^{\alpha\dot{\alpha}}\psi_{\alpha}\bar{\partial}_{\dot{\alpha}}\bar{\lambda}+\frac{1}{2}e^{\alpha\dot{\alpha}}\bar{\psi}_{\dot{\alpha}}\partial_{\alpha}\lambda.|_{Y=0}\label{Real_eq_1}\\
 & DF+ie\partial\bar{\partial}F-\bar{\psi\partial}\lambda=0.\\
 & D\bar{F}+ie\partial\bar{\partial}\bar{F}-\psi\partial\bar{\lambda}=0.\\
 & D\lambda+ie\partial\bar{\partial}\lambda+i\psi\partial F=0.\\
 & D\bar{\lambda}+ie\partial\bar{\partial}\bar{\lambda}+i\bar{\psi\partial}\bar{F}=0.\label{Real_eq_5}
\end{align}

To go off-shell one has to switch on external currents for $F$ and
$\lambda$. In principle, one could start with off-shell systems for
$F$ and $\lambda$ presented in \eqref{vec_eq_off_1}-\eqref{vec_eq_off_2},
\eqref{Dirac_off_eq} and then look for consistent supersymmetric
$\psi$-corrections. However, there is a more efficient way along
the lines of \citep{misuna_1}: to make use of a supersymmetric generalization
of an electric current, which is provided by a linear multiplet. This
includes conserved vector, unconstrained Majorana spinor and unconstrained
pseudoscalar. So one can first find an unfolded system for a linear
multiplet and then couple it to \eqref{Real_eq_1}-\eqref{Real_eq_5}.

To this end one takes unfolded systems for an electric current $J$,
off-shell spinor $(\chi,\bar{\chi})$ (it is convenient to separate
a Majorana spinor into two conjugate Weyl's) and off-shell scalar
$C$, and add possible consistent terms with $\psi$ and $\bar{\psi}$
which mix component fields. This results in
\begin{align}
 & DJ+ie\partial\bar{\partial}J+iey\bar{y}\frac{\varsigma(\varsigma+3)\partial_{p}^{2}}{(N+1)(N+2)(\bar{N}+1)(\bar{N}+2)}J+\nonumber \\
 & +ey\bar{\partial}\frac{1}{(N+1)(N+2)}(J^{-}+2\partial_{p}^{2}J^{0})+e\partial\bar{y}\frac{1}{(\bar{N}+1)(\bar{N}+2)}(J^{+}+2\partial_{p}^{2}J^{0})+\nonumber \\
 & +(\psi\partial\Pi^{+}+2\psi\partial\partial_{p}\Pi^{-}-2i\psi y\frac{\partial_{p}^{2}}{(\varsigma+\frac{3}{2})}\Pi^{+}-i\psi y\frac{(\varsigma+\frac{5}{2})\partial_{p}}{(\varsigma+\frac{1}{2})(\varsigma+\frac{3}{2})}\Pi^{-})\chi+\nonumber \\
 & +(\bar{\psi\partial}\Pi^{-}+2\bar{\psi\partial}\partial_{p}\Pi^{+}-2i\bar{\psi y}\frac{\partial_{p}^{2}}{(\varsigma+\frac{3}{2})}\Pi^{-}-i\bar{\psi y}\frac{(\varsigma+\frac{5}{2})\partial_{p}}{(\varsigma+\frac{1}{2})(\varsigma+\frac{3}{2})}\Pi^{+})\bar{\chi}=0.\label{lin_eq_1}\\
 & D\chi+ie\partial\bar{\partial}\chi+iey\bar{y}\frac{\partial_{p}^{2}}{(\varsigma+\frac{3}{2})^{2}}\chi+ey\bar{\partial}\frac{\partial_{p}}{(\varsigma+\frac{1}{2})(\varsigma+\frac{3}{2})}\Pi^{-}\chi+e\partial\bar{y}\frac{\partial_{p}}{(\varsigma+\frac{1}{2})(\varsigma+\frac{3}{2})}\Pi^{+}\chi-\nonumber \\
 & -\frac{i}{2}\bar{\psi\partial}J^{0}+\frac{1}{2}\bar{\psi y}\frac{1}{(\varsigma+1)}J^{+}+\frac{1}{2}\bar{\psi y}\frac{\varsigma}{(\varsigma+1)(\varsigma+2)}J^{0}-\frac{i}{2}\bar{\psi\partial}p\frac{1}{p\partial_{p}+1}J^{-}-\frac{1}{2}\bar{\psi\partial}C+\frac{1}{2}\bar{\psi y}\frac{\partial_{p}}{(\varsigma+1)}C=0.\\
 & D\bar{\chi}+ie\partial\bar{\partial}\bar{\chi}+iey\bar{y}\frac{\partial_{p}^{2}}{(\varsigma+\frac{3}{2})^{2}}\bar{\chi}+e\partial\bar{y}\frac{\partial_{p}}{(\varsigma+\frac{1}{2})(\varsigma+\frac{3}{2})}\Pi^{+}\bar{\chi}+ey\bar{\partial}\frac{\partial_{p}}{(\varsigma+\frac{1}{2})(\varsigma+\frac{3}{2})}\Pi^{-}\bar{\chi}-\nonumber \\
 & -\frac{i}{2}\psi\partial J^{0}+\frac{1}{2}\psi y\frac{1}{(\varsigma+1)}J^{-}+\frac{1}{2}\psi y\frac{\varsigma}{(\varsigma+1)(\varsigma+2)}J^{0}-\frac{i}{2}\psi\partial p\frac{1}{p\partial_{p}+1}J^{+}+\frac{1}{2}\psi\partial C-\frac{1}{2}\psi y\frac{\partial_{p}}{(\varsigma+1)}C=0.
\end{align}
\begin{align}
 & DC+ie\partial\bar{\partial}C+iey\bar{y}\frac{\partial_{p}^{2}}{(\varsigma+1)(\varsigma+2)}C+i\psi\partial\Pi^{+}\chi-i\bar{\psi\partial}\Pi^{-}\bar{\chi}-\psi y\frac{\partial_{p}}{(\varsigma+\frac{3}{2})}\Pi^{-}\chi+\bar{\psi y}\frac{\partial_{p}}{(\varsigma+\frac{3}{2})}\Pi^{+}\bar{\chi}=0,\label{lin_eq_4}
\end{align}
which is an unfolded form of the linear multiplet. The pseudoscalar
nature of $C$ manifests in opposite signs between terms with $\psi$
and $\bar{\psi}$, which mix it with the spinor.

Now coupling of \eqref{lin_eq_1}-\eqref{lin_eq_4} to \eqref{Real_eq_1}-\eqref{Real_eq_5}
yields
\begin{align}
 & \mathrm{d}A=\frac{i}{4}E\partial\partial F+\frac{i}{4}\bar{E\partial\partial}\bar{F}+\frac{1}{2}e^{\alpha\dot{\alpha}}\psi_{\alpha}\bar{\partial}_{\dot{\alpha}}\bar{\lambda}+\frac{1}{2}e^{\alpha\dot{\alpha}}\bar{\psi}_{\dot{\alpha}}\partial_{\alpha}\lambda.|_{Y=0}\label{real_eq_off_1}\\
 & DF+ie\partial\bar{\partial}F+iey\bar{y}\frac{1}{(\varsigma+1)(\varsigma+2)}J^{+}+ey\bar{\partial}\frac{2}{(\varsigma+1)(\varsigma+2)}J^{0}-\bar{\psi\partial}\lambda-\psi y\frac{2i}{(\varsigma+\frac{3}{2})}\Pi^{+}\chi=0.\\
 & D\bar{F}+ie\partial\bar{\partial}\bar{F}+iey\bar{y}\frac{1}{(\varsigma+1)(\varsigma+2)}J^{-}+e\partial\bar{y}\frac{2}{(\varsigma+1)(\varsigma+2)}J^{0}-\psi\partial\bar{\lambda}-\bar{\psi y}\frac{2i}{(\varsigma+\frac{3}{2})}\Pi^{-}\bar{\chi}=0.\\
 & D\lambda+ie\partial\bar{\partial}\lambda-ey\bar{\partial}\frac{2}{(\varsigma+\frac{1}{2})(\varsigma+\frac{3}{2})}\Pi^{-}\bar{\chi}-ey\bar{y}\frac{2i\partial_{p}}{(\varsigma+\frac{3}{2})^{2}}\Pi^{+}\bar{\chi}+\\
 & +i\psi\partial F+i\psi y\frac{1}{(\varsigma+1)}C-\psi y\frac{\varsigma}{(\varsigma+1)(\varsigma+2)}J^{0}=0.\\
 & D\bar{\lambda}+ie\partial\bar{\partial}\bar{\lambda}-e\partial\bar{y}\frac{2}{(\varsigma+\frac{1}{2})(\varsigma+\frac{3}{2})}\Pi^{+}\chi-ey\bar{y}\frac{2i\partial_{p}}{(\varsigma+\frac{3}{2})^{2}}\Pi^{-}\chi+\\
 & +i\bar{\psi\partial}\bar{F}-i\bar{\psi y}\frac{1}{(\varsigma+1)}C-\bar{\psi y}\frac{\varsigma}{(\varsigma+1)(\varsigma+2)}J^{0}=0.\label{real_eq_off_5}
\end{align}
Together equations \eqref{lin_eq_1}-\eqref{lin_eq_4}, \eqref{real_eq_off_1}-\eqref{real_eq_off_5}
form an unfolded off-shell system for the vector supermultiplet. Let
us stress a characteristic feature of the used approach to build off-shell
supersymmetric models: in a conventional Lagrangian formulation the
off-shell vector supermultiplet includes the Maxwell field, the gaugino
and the auxiliary pseudoscalar which vanishes on-shell, while in the
unfolded approach there is an infinite number of descendant fields,
in particular those which vanish on-shell (these are exactly the unfolded
module of the linear multiplet). What is their relation to the conventional
off-shell vector supermultiplet with only one auxiliary pseudoscalar?
The answer is that among plenty unfolded fields presented in \eqref{lin_eq_1}-\eqref{real_eq_off_5},
the only primaries when considering Minkowski space with $\psi=\bar{\psi}=0$
are $A$, $\lambda_{\alpha}(p=0)$, $\bar{\lambda}_{\dot{\alpha}}(p=0)$
and $C(p=0)$ -- which precisely corresponds to the field content
of the Lagrangian off-shell vector supermultiplet. But in our construction
this pseudoscalar appears as a part of the external current for the
vector supermultiplet, necessary for relaxing on-shell constraints.

Analogously to what is done in Section \ref{SEC_WZ} for the Wess--Zumino
model, it is possible to recombine component fields of the off-shell
vector supermultiplet by modification their $p$-dependence, such
that the unfolded system gets simplified. All spinor modules $\lambda$,
$\bar{\lambda}$, $\chi$, $\bar{\chi}$ can be combined into a single
off-shell Majorana module $\Lambda$, while $J^{+}$ and $J^{-}$
are naturally included to $F$ and $\bar{F}$ as their $p^{2}$-dependent
parts and $J^{0}$ and $C$ get united into a single module $\Phi$
as $p$-even and $p$-odd parts respectively

\begin{align}
 & \Phi=p^{2}J^{0}+pC,\\
 & \Phi(Y|p|x)=\sum_{M=1,n=0}^{\infty}\frac{1}{M!(n!)^{2}}\Phi_{\alpha(n),\dot{\alpha}(n)}^{(M)}p^{M}(y^{\alpha}\bar{y}^{\dot{\alpha}})^{n},
\end{align}
with a constraint on the eigenvalues $\lambda_{\varsigma}$ of $\varsigma$
on the subspace $\Pi^{e}\Phi$
\begin{equation}
\lambda_{\varsigma}|_{\Pi^{e}\Phi}\geq1,
\end{equation}
which encodes conservation of the electric current ($J^{0}$ has zero
divergence and hence no scalar descendants). Another point one has
to ensure is pseudo-reality of $C$. This can be elegantly built into
a modified conjugation operation $h.c.$, which originally just exchanges
dotted and undotted spinors in the unfolded equations. The modified
conjugation now also flips the sign of $p$ in $\Phi$, thus multiplying
$C$ by $-1$ 
\begin{equation}
h.c.:(\alpha,\dot{\beta})\rightarrow(\dot{\alpha},\beta),\qquad\Phi(p)\rightarrow\Phi(-p).
\end{equation}
As a result, the unfolded off-shell system for the vector supermultiplet
now reads
\begin{align}
 & \mathrm{d}A=\frac{i}{4}E\partial\partial F+\frac{i}{4}\bar{E\partial\partial}\bar{F}+\frac{1}{2}e^{\alpha\dot{\alpha}}\psi_{\alpha}\bar{\partial}_{\dot{\alpha}}\Lambda+\frac{1}{2}e^{\alpha\dot{\alpha}}\bar{\psi}_{\dot{\alpha}}\partial_{\alpha}\Lambda.|_{p,Y=0}\\
 & DF+ie\partial\bar{\partial}F+iey\bar{y}\frac{\partial_{p}^{2}}{(\varsigma+1)(\varsigma+2)}F+ey\bar{\partial}\frac{2\partial_{p}^{2}\Pi^{e}}{(\varsigma+1)(\varsigma+2)}\Phi-\bar{\psi\partial}\Pi^{e}\Lambda^{+}+i\psi y\frac{\partial_{p}}{(\varsigma+\frac{3}{2})}\Pi^{o}\Lambda^{+}=0.\\
 & D\bar{F}+ie\partial\bar{\partial}\bar{F}+iey\bar{y}\frac{\partial_{p}^{2}}{(\varsigma+1)(\varsigma+2)}\bar{F}+e\partial\bar{y}\frac{2\partial_{p}^{2}\Pi^{e}}{(\varsigma+1)(\varsigma+2)}\Phi-\psi\partial\Pi^{e}\Lambda^{-}+i\bar{\psi y}\frac{\partial_{p}}{(\varsigma+\frac{3}{2})}\Pi^{o}\Lambda^{-}=0.\\
 & D\Lambda+ie\partial\bar{\partial}\Lambda+iey\bar{y}\frac{\partial_{p}^{2}}{(\varsigma+\frac{3}{2})^{2}}\Lambda+ey\bar{\partial}\frac{\partial_{p}}{(\varsigma+\frac{1}{2})(\varsigma+\frac{3}{2})}\Lambda^{-}+e\partial\bar{y}\frac{\partial_{p}}{(\varsigma+\frac{1}{2})(\varsigma+\frac{3}{2})}\Lambda^{+}+\nonumber \\
 & +\left(i\psi\partial F+i\psi\partial(\partial_{p}\Pi^{e}+\Pi^{o})\Phi-\psi y\frac{\partial_{p}}{(\varsigma+1)}\bar{F}-\psi y\frac{\partial_{p}}{(\varsigma+1)}(\frac{\varsigma}{(\varsigma+2)}\partial_{p}\Pi^{e}-\Pi^{o})\Phi+h.c.\right)=0.\\
 & D\Phi+ie\partial\bar{\partial}\Phi+iey\bar{y}\frac{\partial_{p}^{2}}{(\varsigma+1)(\varsigma+2)}(1-\frac{2\Pi^{e}}{(\varsigma+1)(\varsigma+2)})\Phi+ey\bar{\partial}\frac{1}{\varsigma(\varsigma+1)}\bar{F}+e\partial\bar{y}\frac{1}{\varsigma(\varsigma+1)}F+\nonumber \\
 & +\left(\frac{i}{2}\psi y\frac{\partial_{p}}{(\varsigma+\frac{3}{2})}(\frac{(\varsigma+\frac{5}{2})}{(\varsigma+\frac{1}{2})}+\partial_{p})\Pi^{e}\Lambda^{-}-\frac{1}{2}\psi\partial p\frac{1}{p\partial_{p}+1}(1-\partial_{p})\Pi^{o}\Lambda^{+}+h.c.\right)=0.
\end{align}
Although being somewhat bulky, it obviously looks much simpler than
its ``non-compressed'' form \eqref{lin_eq_1}-\eqref{real_eq_off_5}.

\section{Conclusion\label{SEC_CONCLUSION}}

In the paper we have constructed and analyzed unfolded off-shell systems
for the chiral and vector supermultiplets, by formulating corresponding
on-shell systems and coupling them to the external currents. We worked
in the multispinor formalism and found that our formulation reveals
certain new interesting features.

Analyzing the unfolded off-shell Wess--Zumino model, we have discovered
a way to reorganize unfolded fields in a more compact way, with only
one scalar and one spinor unfolded modules, which seems elusive in
the original tensorial formulation of \citep{susnf0,susnf1}. The
clue is the use of an additional variable $p$, which was originally
introduced as a formal parameter cataloging off-shell descendants
of the primary fields, resulting from the action of kinetic operators
\citep{misuna_1}. Comparing to the tensorial formulation of \citep{susnf1},
this $p$ accounts for the traces of the off-shell traceful unfolded
fields. However, in supersymmetric model it allows one to unite dynamical
and auxiliary scalars of Wess--Zumino model into a single unfolded
module. Moreover, it turns out that in this new form unfolded off-shell
equations become pseudo-real, looking completely identical for chiral
and anti-chiral supermultiplets. And an additional constraint which
chooses between chiral and anti-chiral systems takes a form of a simple
relation constraining $p$-dependence of the unfolded spinor module.
At this moment $p$ becomes not just a book-keeping parameter but
an actual ``alive'' variable. It cannot be interpreted as just a
counterpart of tensor traces in \citep{susnf1} anymore (in fact,
it becomes hard to find any simple straightforward interpretation
for it in tensorial terms). It must be emphasized that this variable
is not specially designed and introduced in order to merge two scalar
modules into one (otherwise, this would be a triviality): it necessarily
appears for all unfolded off-shell relativistic fields written in
terms of multispinors \citep{misuna_1}.

Further, we have constructed an unfolded description of the vector
supermultiplet along the lines of \citep{misuna_1}. To this end we
first found an unfolded on-shell system, then built an unfolded formulation
for the linear multiplet which is an external source for the vector
supermultiplet, and finally coupled them together. Once again, by
modifying $p$-dependence we managed to put the system in a more concise
form, analogously to what happened for the Wess--Zumino model. And
if in that case there were additional (anti-)chirality constraints
on the unfolded modules, formulated in terms of $p$-dependence, for
the vector supermultiplet such additional constraints are electric
current conservation and pseudoscalar parity condition, which are
also formulated in terms of $p-Y$.

Our results show that the unfolded formalism is a fruitful instrument
for analyzing supersymmetric theories, which can reveal new features
even for the well-studied SUSY models, like those considered in the
paper. It would be interesting to apply the proposed approach to more
complicated theories, including higher-spin supermultiplets, extended-SUSY
and interacting models. An important property of the unfolded formalism
is that it allows one to freely pass between different base space-time
manifolds (this is possible when the unfolded consistency condition
holds regardless of the dimension of the base manifold, which is usually
the case), which gives rise to various dualities \citep{vashol}.
In particular, one can directly uplift unfolded SUSY-models, formulated
in component terms in Minkowski space, to superspace formulation \citep{susnf0,susnf1}.
It would be interesting to see how the superspace unfolded generalizations
of the space-time formulations, constructed in the paper, fit the
standard superspace formulations (in particular, what are relation
between the spectra of auxiliary fields).

\section*{Acknowledgments}

The author is grateful to Karapet Mkrtchyan for pointing out a typo.
The research was supported by the Russian Basic Research Foundation
Grant \textnumero{} 20-02-00208 and by the Alexander von Humboldt
Foundation.


\begin{thebibliography}{99}
\bibitem{unf1}M.A. Vasiliev \emph{Yad.Fiz.} \textbf{32} (1980) 855-861,
\emph{Sov.J.Nucl.Phys.} \textbf{32} (1980) 439.

\bibitem{unf2}M.A. Vasiliev, \emph{Annals Phys.} \textbf{190} (1989)
59-106.

\bibitem{unf4}M.A. Vasiliev, \emph{Int.J.Geom.Meth.Mod.Phys.} \textbf{3}
(2006) 37-80 \href{https://arxiv.org/abs/hep-th/0504090}{[hep-th/0504090]}.

\bibitem{vaseq1}M.A. Vasiliev, \emph{Phys.Lett.B} \textbf{243} (1990)
378-382.

\bibitem{vaseq2}M.A. Vasiliev, \emph{Phys.Lett.B} \textbf{285} (1992)
225-234.

\bibitem{susnf0}D.S. Ponomarev, M.A. Vasiliev, \emph{JHEP} \textbf{1201}
(2012) 152 \href{https://arxiv.org/abs/1012.2903}{[arXiv:1012.2903]}.

\bibitem{susnf1}N.G. Misuna, M.A. Vasiliev, \emph{JHEP} \textbf{05}
(2014) 140 \href{https://arxiv.org/abs/1301.2230}{[arXiv:1301.2230]}.

\bibitem{susnf2}M.V. Khabarov, Yu.M. Zinoviev, \emph{Nucl.Phys.B}
\textbf{953} (2020) 114959 \href{https://arxiv.org/abs/2001.07903}{[arXiv:2001.07903]}.

\bibitem{susnf3}I.L. Buchbinder, T.V. Snegirev, Yu. M. Zinoviev,
\emph{JHEP} \textbf{08} (2016) 075 \href{https://arxiv.org/abs/1606.02475}{[arXiv:1606.02475]}.

\bibitem{misuna_1}N.G. Misuna, \emph{Phys.Lett.B} \textbf{798} (2019)
134956 \href{https://arxiv.org/abs/1905.06925}{[arXiv:1905.06925]}.

\bibitem{off1}O.V. Shaynkman, M.A. Vasiliev, \emph{Theor.Math.Phys.}
\textbf{123} (2000) 683-700, \emph{Teor.Mat.Fiz.} \textbf{123} (2000)
323-344 \href{https://arxiv.org/abs/hep-th/0003123}{[hep-th/0003123]}.

\bibitem{off2}A. Sagnotti, E. Sezgin, P. Sundell, \emph{On higher
spins with a strong Sp(2,R) condition}, contribution to 1st Solvay
Workshop on Higher Spin Gauge Theories, 100-131, \href{https://arxiv.org/abs/hep-th/0501156}{[hep-th/0501156]}.

\bibitem{off3}X. Bekaert, S. Cnockaert, C. Iazeolla, M.A. Vasiliev,
\emph{Nonlinear higher spin theories in various dimensions}, contribution
to 1st Solvay Workshop on Higher Spin Gauge Theories, 132-197, \href{https://arxiv.org/abs/hep-th/0503128}{[hep-th/0503128]}.

\bibitem{misuna_2}N.G. Misuna, \emph{JHEP} \textbf{12} (2021) 172
\href{https://arxiv.org/abs/2012.06570}{[arXiv:2012.06570]}.

\bibitem{vashol}M.A. Vasiliev, \emph{J.Phys.A} \textbf{46} (2013)
214013 \href{https://arxiv.org/abs/1203.5554}{[arXiv:1203.5554]}.
\end{thebibliography}
\end{document}